\documentclass[11pt]{article}

\input epsf
\usepackage{latexsym}
\usepackage{amssymb}
\usepackage{amsmath}
\usepackage[dvips]{graphicx}
\usepackage{array}

\begin{document}
\centerline{\Large \bf  Completing magnetic field generation from  gravitationally coupled}
\vskip 0.3cm
\centerline{\Large \bf  electrodynamics with the curvaton mechanism}

\vskip 2 cm

\centerline{Kerstin E. Kunze
\footnote{E-mail: kkunze@usal.es} }

\vskip 0.3cm

\centerline{{\sl Departamento de F\'\i sica Fundamental} and {\sl IUFFyM},}
\centerline{{\sl Universidad de Salamanca,}}
\centerline{{\sl Plaza de la Merced s/n, 37008 Salamanca, Spain }}

\vskip 1.5cm

\centerline{\bf Abstract}
\vskip 0.5cm
\noindent
Primordial magnetic fields of cosmologically interesting field strengths can be generated from gravitationally coupled electrodynamics during inflation.
As the cosmological constraints require this to be power law inflation it is not possible to generate at the same time the curvature perturbation from inflation.
Therefore here a completion is considered whereby the large scale magnetic field is generated during inflation and the primordial 
curvature mode in a subsequent era from a curvaton field. It is found that constraints on the model to obtain strong magnetic fields and those to suppress the amplitude of the curvature perturbation generated during inflation can be simultaneously satisfied for magnetic seed fields $B_s\stackrel{>}{\sim}10^{-30}$ G.

\vskip 1cm

\section{Introduction}
\setcounter{equation}{0}
Observations indicate the existence of large scale magnetic fields in the universe. Evidence for magnetic fields at galactic and cluster scale 
has been obtained for decades with field strengths of order $10^{-6}$ G \cite{obs-gal}. However, over recent years several groups have found indications of truly cosmologically magnetic fields, pervading space with intensities of order $10^{-15}$ G \cite{vovk}. There is a multitude of proposals to explain the existence 
of cosmological magnetic fields (for recent reviews, see, e.g., \cite{reviews}). Placing the generation mechanism in the very early universe during inflation by amplifying significantly perturbations in the electromagnetic field requires to break conformal invariance in flat backgrounds \cite{tw} which however is not the case in open geometries \cite{open} (see, however, \cite{ard}). 

The conformal invariance of Maxwell's equations in four dimensions is broken in models in which the electromagnetic field is gravitationally coupled.
Couplings between curvature terms and the Maxwell tensor in the Lagrangian of the form $RF_{\mu\nu}F^{\mu\nu}$, $R_{\mu\nu}F^{\mu\kappa}F^{\nu}_{\;\;\;\kappa}$ and $R_{\mu\nu\lambda\kappa}F^{\mu\nu}F^{\lambda\kappa}$ 
are present when describing, e.g., the propagation of a photon in a curved background \cite{dh}. In Fourier space the resulting mode equation for the gauge potential of the electromagnetic field in an expanding background includes a term which during inflation can cause sufficient amplification on superhorizon scales so as to generate a strong enough magnetic field in order to  seed the galactic dynamo. This was first proposed in \cite{tw} and investigated in more detail in \cite{kk}.

\begin{figure}[h!]
\centerline{\epsfxsize=3.3in\epsfbox{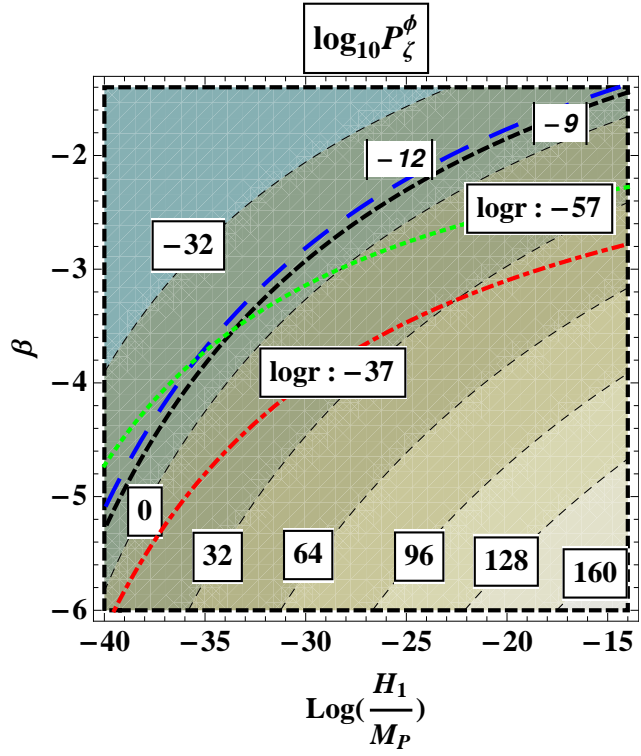}}
\caption{Contour plot of  $\log_{10}{\cal P}^{\phi}_{\zeta}$  shown for 
$k_p=0.002$ Mpc$^{-1}$. The contour lines corresponding to ${\cal P}^{\phi}_{\zeta}=10^{-9}$ (thick, black dashed line) and ${\cal P}^{\phi}_{\zeta}=10^{-12}$
(thick, blue long dashed line) are marked. The red, dot-dashed and green, dotted lines show the maximal value of the logarithm of the ratio of magnetic to background radiation energy density of the model \cite{kk}, corresponding to $r=10^{-37}$ (red, dot-dashed) and $r=10^{-57}$ (green, dotted).}
\label{fig1}
\end{figure}

In \cite{kk} the final spectrum of the magnetic field is calculated assuming the inflationary stage is directly matched to 
the standard radiation dominated era at some conformal time $\tau=\tau_1$, so that the scale factor has the form 
\begin{eqnarray}
a(\tau)=\left\{
\begin{array}{lr}
a_1\left(\frac{\tau}{\tau_1}\right)^{\beta} & \tau<\tau_1\\
a_1\left(\frac{\tau-2\tau_1}{-\tau_1}\right) & \tau\geq\tau_1
\end{array}
\right.
\end{eqnarray}
and the line element has the form $ds^2=a^2(\tau)(-d\tau^2+dx^2+dy^2+dz^2)$. 
In the following $a_1\equiv 1$. For $\tau<\tau_1$, de Sitter inflation takes place for $\beta=-1$ and power law inflation for 
$-\infty<\beta<-1$.
In \cite{kk} the fractional magnetic field energy density, $r=\frac{\rho_B}{\rho_{\gamma}}$ on a galactic scale of order 1 Mpc has been found as
\begin{eqnarray}
r(\omega_G)=10^{-79+52\nu}\left[\Gamma(\nu)\right]^2\left(\frac{1}{2}-\nu\right)^2\left(\frac{\xi_2}{4}\right)^{-\nu}\left(\frac{H_1}{M_P}\right)^{\nu+\frac{1}{2}},
\end{eqnarray}
where $\nu\equiv|\beta+\frac{3}{2}|$, $\xi_2=\frac{10\beta-7}{7\beta-10}$ and  $H_1$ is the Hubble parameter at the beginning of the radiation dominated era, at $\tau=\tau_1$ which determines the reheat temperature.
It was found that a magnetic field strength $B>10^{-20}$ G corresponding to $r>10^{-37}$ can be reached for $\beta<-2.8$ and $\frac{H_1}{M_P}<10^{-18}$, depending on $\beta$ (cf. figure \ref{fig1}).
The constraint on $\beta$ prevents that the observed curvature perturbation is generated during inflation since the resulting 
spectral index is too small.  In slow roll inflation power law inflation \cite{lm}  is realized by an exponential potential of the inflaton $\phi$ of the form (e.g, \cite{pu})
\begin{eqnarray}
V(\phi)=V_i\exp\left[4\sqrt{\frac{\pi}{p}}\frac{\left(\phi-\phi_i\right)}{M_P}\right],
\end{eqnarray}
where the index $i$ indicates initial values and $p$ determines the scale factor in cosmic time $dt=ad\tau$, that is $a\sim t^p$. Thus $p=\frac{\beta}{\beta+1}$. 
The slow roll parameters are given by $\epsilon=\frac{\eta}{2}=\frac{1}{p}$. Ending inflation by bubble formation puts the constraint $p<10$ \cite{dl,ll}.
$p$ as a function of $\beta$ is monotonically growing for $\beta<0$, approaching 1 for $\beta\rightarrow-\infty$ and at $\beta=-2.8$, $p=1.56$, so that the upper limit on $p$ does not put any additional constraint on the model at hand. The spectral index of the curvature perturbation created during inflation, $n_s=1+2\eta-6\epsilon$, is in the range $-1$ and $-0.29$ for $\beta<-2.8$.
The power spectrum of the curvature perturbation evaluated at 
horizon crossing is given by \cite{pu}
\begin{eqnarray}
{\cal P}^{\phi}_{\zeta}=\frac{1}{\pi M_P^2}\left[2^{\mu-\frac{3}{2}}\frac{\Gamma(\mu)}{\Gamma(\frac{3}{2})}\right]^2
\left(\mu-\frac{1}{2}\right)^{-2\mu+1}\left(\frac{H^2}{\epsilon}\right)_{k=aH},
\label{pzph}
\end{eqnarray}
where $\mu\equiv\frac{3}{2}+\frac{1}{p-1}$.
The amplitude of the curvature perturbation generated during inflation should only contribute negligibly to the final curvature perturbation.
Starting with the slow roll equations 
\begin{eqnarray}
3H\dot{\phi}+V_{\phi}&=&0\\
H^2&=&\frac{8\pi}{3M_P^2}V
\end{eqnarray}
the evolution of the inflaton $\phi$ is found to be
\begin{eqnarray}
\frac{\phi-\phi_i}{M_P}=-\frac{1}{2}\sqrt{\frac{p}{\pi}}(\beta+1)\ln\left(\frac{\tau}{\tau_i}\right).
\end{eqnarray}
Thus the potential is given by 
\begin{eqnarray}
\frac{V(\phi)}{M_P^4}=\frac{3\beta^2}{8\pi}\left(\frac{\tau_1}{M_P^{-1}}\right)^{-2}\left(\frac{\tau}{\tau_1}\right)^{-2(\beta+1)}.
\end{eqnarray}
This allows to calculate $H_k\equiv H(\phi_k)=H|_{k=aH}$. Using that $aH=\frac{\beta}{\tau}$ and $a_1=a(\tau_1)=1$ it follows that 
\begin{eqnarray}
\left(\frac{H_k}{M_P}\right)^2=\left(\frac{H_1}{M_P}\right)^{-2\beta}\left(\frac{k}{M_P}\right)^{2(\beta+1)}.
\end{eqnarray}
So that finally, using $\Omega_{\gamma,0}=\left(\frac{H_1}{H_0}\right)^2\left(\frac{a_1}{a_0}\right)^4$,
\begin{eqnarray}
\left(\frac{H_k}{M_P}\right)^2=(5.24\times 10^{-58}\Omega_{\gamma,0}^{-\frac{1}{4}})^{2(\beta+1)}\left(\frac{k_p}{{\rm Mpc}^{-1}}\right)^{2(\beta+1)}
\left(\frac{H_0}{M_P}\right)^{-(\beta+1)}\left(\frac{H_1}{M_P}\right)^{1-\beta},
\label{Hk}
\end{eqnarray}
where $k_p$ is going to be chosen to be the pivot wavenumber of WMAP 7 today, $k_p=0.002$ Mpc$^{-1}$ \cite{wmap7}.
Thus the amplitude of the curvature spectrum at $k_p$ is determined by
\begin{eqnarray}
{\cal P}^{\phi}_{\zeta}&=&\frac{p}{\pi}\left[2^{\mu-\frac{3}{2}}\frac{\Gamma(\mu)}{\Gamma(\frac{3}{2})}\right]^2
\left(\mu-\frac{1}{2}\right)^{-2\mu+1}(5.24\times 10^{-58}\Omega_{\gamma,0}^{-\frac{1}{4}})^{2(\beta+1)}\left(\frac{k_p}{{\rm Mpc}^{-1}}\right)^{2(\beta+1)}
\nonumber\\
&&\times
\left(\frac{H_0}{M_P}\right)^{-(\beta+1)}\left(\frac{H_1}{M_P}\right)^{1-\beta}
\end{eqnarray}
In figure \ref{fig1}  a contour plot of $\log_{10}{\cal P}^{\phi}_{\zeta}$ is shown. 
As can be appreciated from figure \ref{fig1} satisfying the constraints to generate a magnetic field with $B_s\sim 10^{-20}$ G corresponding to $r\sim 10^{-37}$, that is $\beta<-2.8$ leads to curvature perturbations which are in general, too large, not only for the curvaton mechansim to work, but even larger than the observed amplitude ${\cal P}_{\zeta}=2.43\times 10^{-9}$ \cite{wmap7}. 
However, it is possible to generate a magnetic field with $B_s\sim 10^{-30}$ G corresponding to $r\sim 10^{-57}$ which is the lower limit required to seed the galactic dynamo when the non vanishing cosmological constant is taken into account \cite{dt}. In order for the curvature perturbation created during inflation to be subdominant $\beta$ and $\frac{H_1}{M_P}$ have to be in the range, $-4.8\leq\beta\leq -3.6$ and $10^{-40}\leq\frac{H_1}{M_P}\leq 10^{-34}$ which corresponds to reheat temperatures between 0.1 and 100 GeV.
 
\section{Generating the curvature perturbation}
\setcounter{equation}{0}

In the curvaton model the curvature perturbation is generated after inflation has ended, by the curvaton field $\sigma$ which has been subdominant  during inflation \cite{lw}. The spectral index of the final curvature perturbation is given by \cite{dl}
\begin{eqnarray}
n_s=1+2\eta_{\sigma\sigma}-2\epsilon,
\label{ns}
\end{eqnarray}
where 
\begin{eqnarray}
\eta_{\sigma\sigma}\equiv\frac{\bar{M}^2_P}{V}\frac{\partial^2V}{\partial\sigma^2}
\end{eqnarray}
and $\bar{M}_P^2$ is the reduced Planck mass $M_P^2/(8\pi)$.
Assuming the  simplest model \cite{lw,bl} the potential during inflation is defined to be
\begin{eqnarray}
V(\phi,\sigma)=V_i\exp\left[4\sqrt{\frac{\pi}{p}}\frac{\left(\phi-\phi_i\right)}{M_P}\right]+\frac{1}{2}m_{\sigma}^2\sigma^2,
\end{eqnarray}
where the contribution from the curvaton has to be sub leading to the one of the inflaton.
Equation (\ref{ns}) requires $\eta_{\sigma\sigma}$ to be of the order of $\epsilon$ for a nearly scale invariant spectrum.
Thus 
\begin{eqnarray}
\eta_{\sigma\sigma}\simeq\frac{\bar{M}_P^2m_{\sigma}^2}{V_i\exp\left[4\sqrt{\frac{\pi}{p}}\frac{\left(\phi_k-\phi_i\right)}{M_P}\right]}
\simeq
\frac{m^2_{\sigma}}{3H^2_k},
\end{eqnarray}
calculated at the time of first horizon crossing during inflation of the mode $k=aH$.
So that the mass of the curvaton is determined by 
\begin{eqnarray}
\left(\frac{m_{\sigma}}{M_P}\right)=\left[\frac{3}{p}-\frac{3}{2}(1-n_s)\right]^{\frac{1}{2}}\left(\frac{H_k}{M_P}\right).
\end{eqnarray}
\begin{figure}[h!]
\centerline{\epsfxsize=3.1in\epsfbox{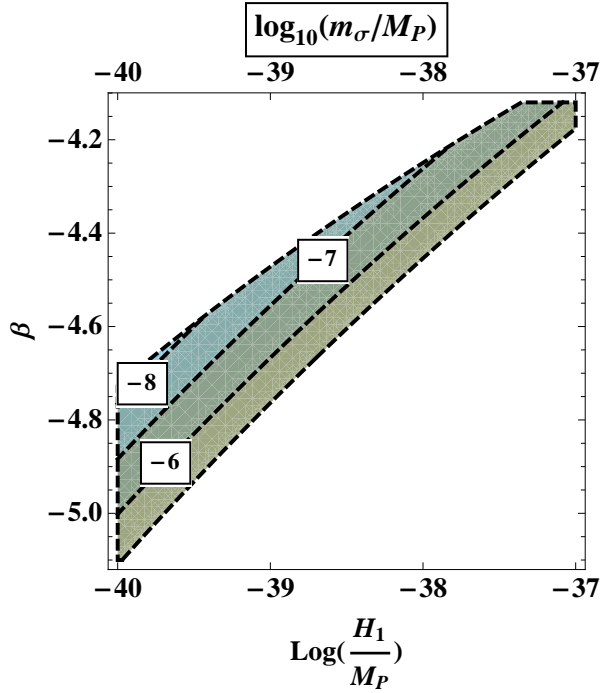}}
\caption{Contour plot of  $\log_{10}\left(\frac{m_{\sigma}}{M_P}\right)$  shown for 
$k_p=0.002$ Mpc$^{-1}$ in the region allowed by the constraints  ${\cal P}^{\phi}_{\zeta}\leq10^{-12}$ and 
 $r>10^{-57}$ for  ${\cal P}_{\zeta}=2.43\times 10^{-9}$ and $1-n_s=0.037$ \cite{wmap7}.}
\label{fig2}
\end{figure}
This is shown in figure \ref{fig2} for the region allowed by the constraints that ${\cal P}^{\phi}_{\zeta}\leq10^{-12}$ and 
 $r>10^{-57}$ for ${\cal P}_{\zeta}=2.43\times 10^{-9}$ and $1-n_s=0.037$ \cite{wmap7}.

The evolution of the background curvaton is determined by 
\begin{eqnarray}
\ddot{\sigma}+3H\dot{\sigma}+V_{\sigma}=0.
\end{eqnarray}
It is assumed \cite{dl} that $|V_{\sigma\sigma}|\ll H^2$ which in the model at hand results in $m_{\sigma}^2t^2\ll 1$ since $p$ is ${\cal O}(1)$ which implies that during and after inflation, before the curvaton becomes effectively massive, $\sigma$ stays approximately constant, $\sigma\simeq\sigma_*$.

In the subsequent analysis we follow \cite{bl}.
The end of inflation takes place at $\tau=\tau_1$ when the Hubble parameter has the value $H_1$. During the following radiation dominated era it evolves as $H^2\sim 1/a^4$ so that 
\begin{eqnarray}
H^2=H_1^2\left(\frac{a_1}{a}\right)^4.
\end{eqnarray}
The curvaton becomes massive at $m_{\sigma}^2=H^2$ so that
\begin{eqnarray}
\left(\frac{a_1}{a_{mass}}\right)^4=\frac{m_{\sigma}^2}{H_1^2}.
\end{eqnarray}
To prevent an additional stage of inflation driven by the curvaton thus requiring the universe to be radiation dominated at
 $\tau_{mass}$ imposes
 $\rho_{rad}(\tau_{mass})\gg\frac{1}{2}m_{\sigma}^2\sigma_*^2$, where $\sigma_*$ is the value of the curvaton during inflation when the modes of the observable perturbations leave the horizon, assuming that, during the late stages of slow roll inflation, the change in $\sigma$ can be neglected.
Then together with the value of $\rho_{rad}$ at the beginning of the radiation dominated era, $\rho_{rad}(\tau_1)=3\bar{M}_P^2H_1^2$, it follows that
\begin{eqnarray}
\sigma_*^2\ll\frac{3 M_P^2}{4\pi}
\end{eqnarray}
which is the same constraint as in the chaotic inflation model of \cite{bl}.

In \cite{lw} two separate cases have been considered. If the curvaton decays during the radiation dominated era, that is until its decay it stays subdominant, the 
resulting curvature perturbation is given by \cite{lw}
\begin{eqnarray}
{\cal P}_{\zeta}=\frac{S^2_{decay}}{16}\frac{H_k^2}{\pi^2\sigma_*^2}
\label{Pzeta1}
\end{eqnarray}
using that $\sigma_k=\sigma_*$ is approximately constant during inflation and $S\equiv\frac{\rho_{\sigma}}{\rho_{rad}}$. Moreover, here $\rho_{\sigma}=\frac{1}{2}m_{\sigma}^2\sigma^2$.
As in \cite{bl} it is found that at the time of decay,
\begin{eqnarray}
S_{decay}=\frac{\sigma_*^2}{6\bar{M}_P^2}\frac{a_{decay}}{a_{mass}}
\end{eqnarray}
and defining the decay constant of the curvaton $\Gamma_{\sigma}$, together with $\Gamma_{\sigma}=H_{decay}$ results in
$a_{decay}/a_{mass}=(m_{\sigma}/\Gamma_{\sigma})^{1/2}$ implying
\begin{eqnarray}
S_{decay}\simeq\frac{\sigma_*^2}{6\bar{M}_P^2}\left(\frac{m_{\sigma}}{\Gamma_{\sigma}}\right)^{\frac{1}{2}}.
\end{eqnarray}
So that finally, the spectrum of the curvature perturbation, for $S_{decay}<1$, is found to be
\begin{eqnarray}
{\cal P}_{\zeta}=\frac{1}{9}\left(\frac{H_k}{M_P}\right)^2\left(\frac{m_{\sigma}}{M_{P}}\right)\left(\frac{\sigma_*}{M_P}\right)^2\left(\frac{\Gamma_{\sigma}}{M_P}\right)^{-1}.
\end{eqnarray}
Moreover, for the perturbations to be Gaussian $H_k/\sigma_*\ll 1$ \cite{lw}. This together with $S_{decay}<1$ determines the range
of possible values of $\sigma_*/H_k$ to be
\begin{eqnarray}
1<\frac{\sigma_*}{H_k}<\frac{1}{4\pi{\cal P}_{\zeta}^{\frac{1}{2}}}.
\end{eqnarray}
Finally, the decay constant $\Gamma_{\sigma}$ is determined by
\begin{eqnarray}
\left(\frac{\Gamma_{\sigma}}{M_P}\right)=\frac{1}{9{\cal P}_{\zeta}}\left(\frac{m_{\sigma}}{M_P}\right)\left(\frac{\sigma_*}{H_k}\right)^2\left(\frac{H_k}{M_P}\right)^4.
\end{eqnarray}
\begin{figure}[h!]
\centerline{\epsfxsize=2.8in\epsfbox{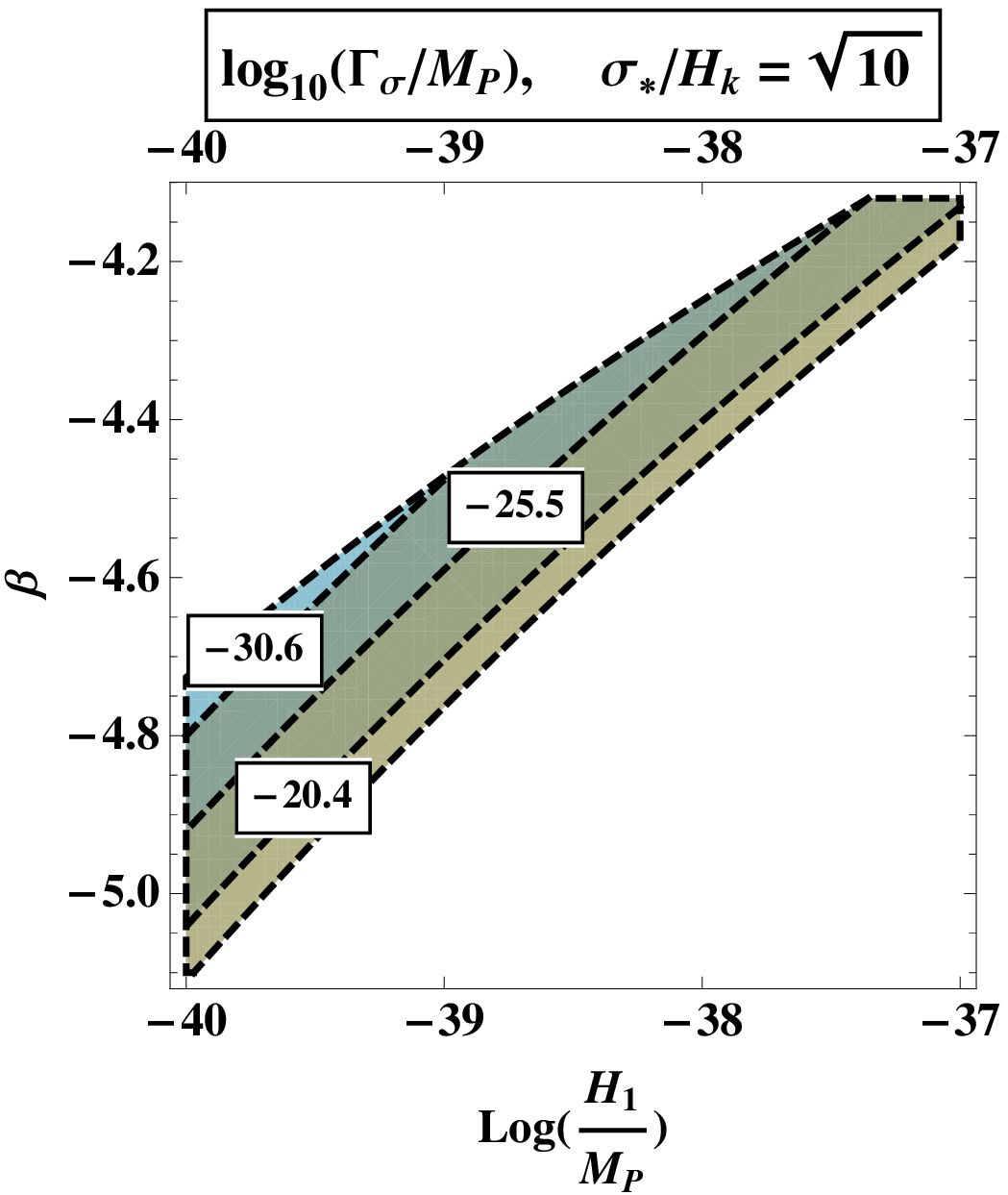}
\hspace{1cm}
\epsfxsize=2.8in\epsfbox{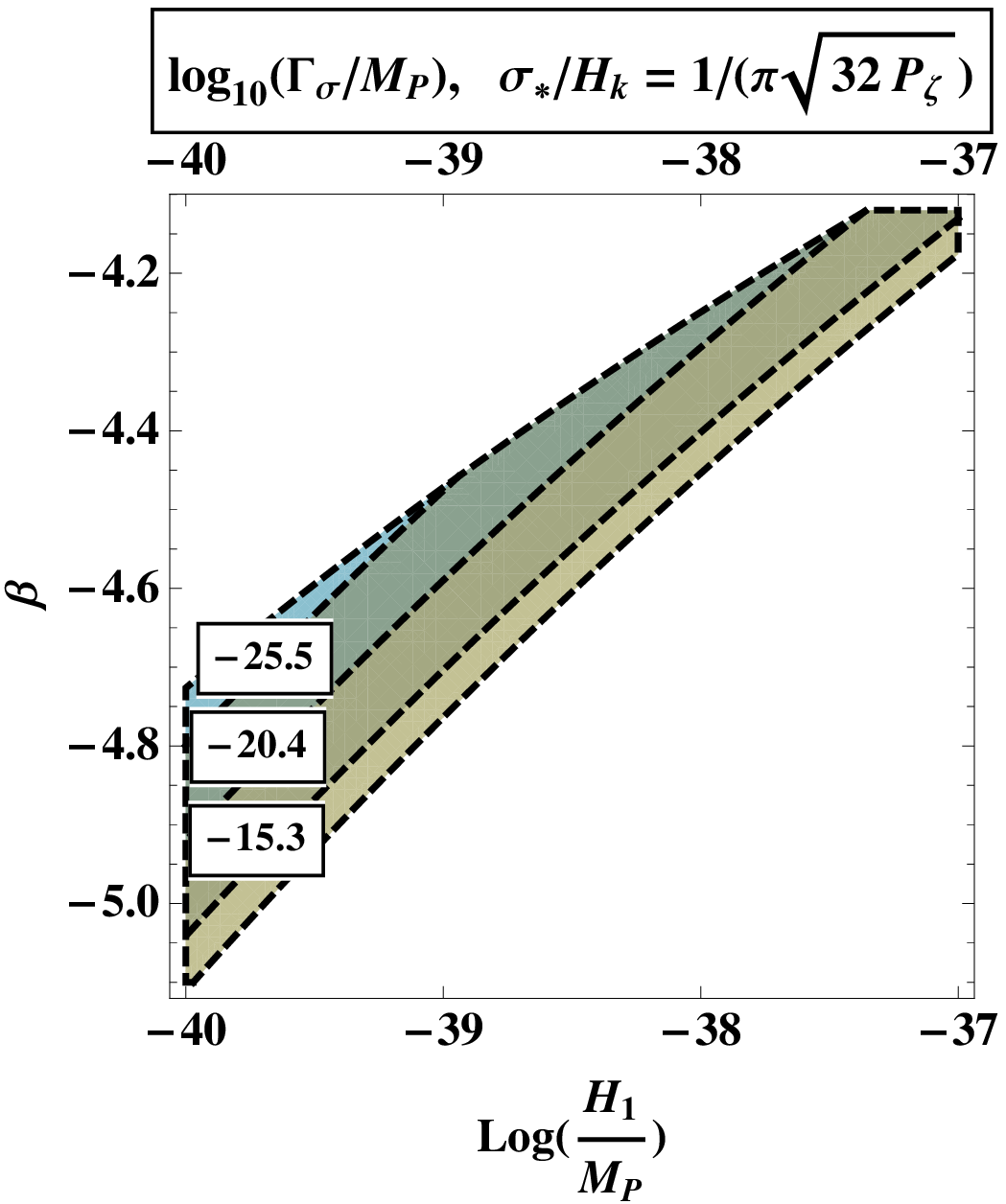}}
\centerline{\epsfxsize=2.8in\epsfbox{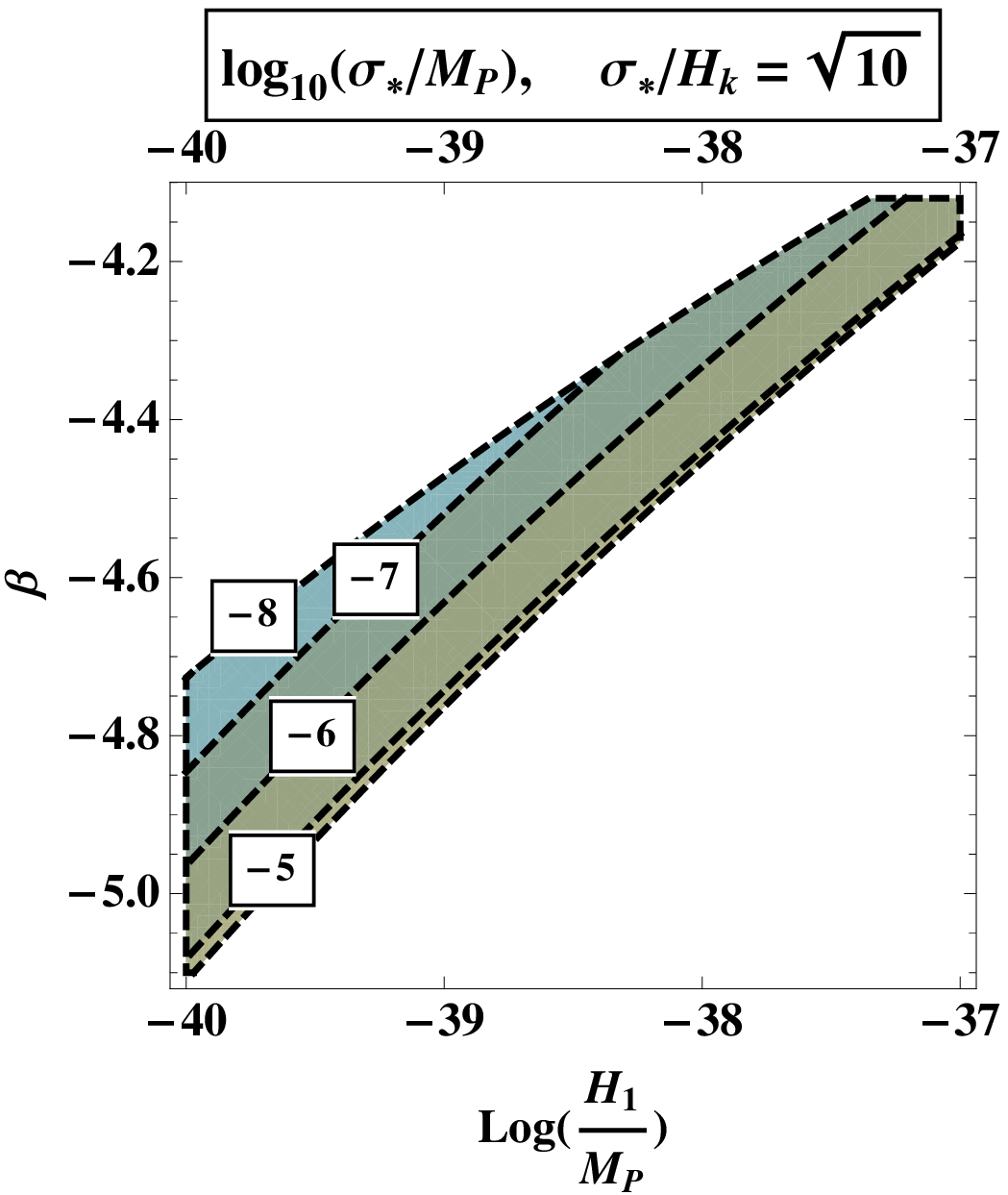}
\hspace{1cm}
\epsfxsize=2.8in\epsfbox{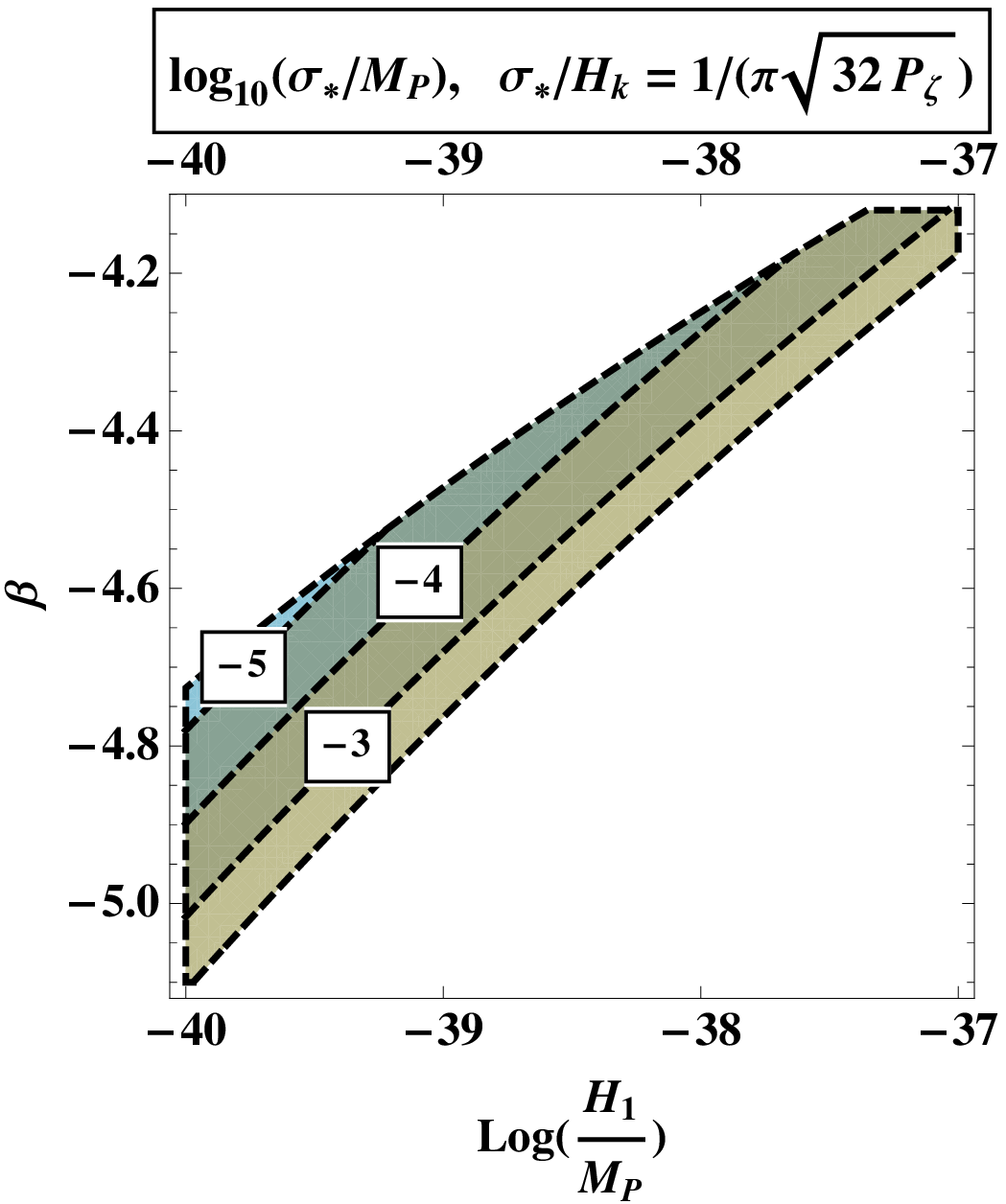}}
\caption{{\it Upper panel:} Contour plot of  $\log_{10}\left(\frac{\Gamma_{\sigma}}{M_P}\right)$.  
{\it Lower panel:} Contour plot of  the corresponding values of $\log_{10}\left(\frac{\sigma_*}{M_P}\right)$. 
All plots are shown for 
$k_p=0.002$ Mpc$^{-1}$ in the region allowed by the constraints  ${\cal P}^{\phi}_{\zeta}\leq10^{-12}$ and 
 $r>10^{-57}$ for different values of $\left(\frac{\sigma_{*}}{H_k}\right)$. 
 The curvature perturbation is assumed to be given by the best fit parameters of WMAP 7, ${\cal P}_{\zeta}=2.43\times 10^{-9}$ and $1-n_s=0.037$ \cite{wmap7}. It is assumed that the curvaton decays during the radiation dominated era, so that $S_{decay}<1$.}
\label{fig3}
\end{figure}
The decay constant $\Gamma_{\sigma}$ and the value of $\sigma_*$ is shown for different values of $\frac{\sigma_*}{H_k}$ in figure \ref{fig3}.
As can be seen $\Gamma_{\sigma}/M_P=H_{decay}>10^{-40}$ which is the minimal value to ensure standard primordial nucleosynthesis.

\begin{figure}[h!]
\centerline{\epsfxsize=3.1in\epsfbox{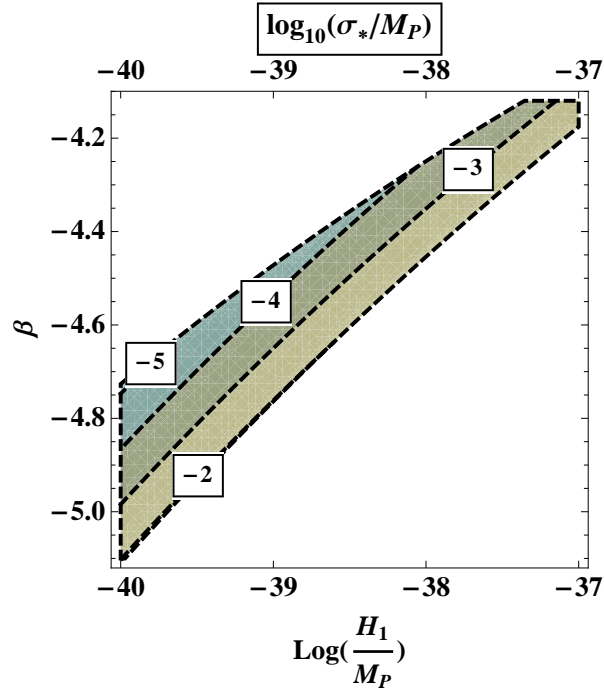}}
\caption{Contour plot of  $\log_{10}\left(\frac{\sigma_*}{M_P}\right)$  shown for 
$k_p=0.002$ Mpc$^{-1}$ in the region allowed by the constraints  ${\cal P}^{\phi}_{\zeta}\leq10^{-12}$ and 
 $r>10^{-57}$ for  ${\cal P}_{\zeta}=2.43\times 10^{-9}$ and $1-n_s=0.037$ \cite{wmap7}. It is assumed that the curvaton dominates before decay, so that $S_{decay}>1$.}
\label{fig4}
\end{figure}

In the opposite case, $S_{decay}>1$, the curvaton dominates before decay and \cite{lw}
\begin{eqnarray}
{\cal P}_{\zeta}\simeq\frac{1}{9}\frac{H_k^2}{\pi^2\sigma_*^2}.
\label{pzeta1}
\end{eqnarray}
In this case the Gaussianity requirement is always satisfied if the curvature perturbation are of the observed magnitude.
Moreover $\sigma_*$ is completely determined by equation (\ref{pzeta1}),
\begin{eqnarray}
\left(\frac{\sigma_*}{M_P}\right)=\frac{1}{3\pi {\cal P}_{\zeta}^{\frac{1}{2}}}\left(\frac{H_k}{M_P}\right)
\end{eqnarray}
which is shown in figure \ref{fig4} for the best fit values of WMAP 7 \cite{wmap7}.
In this case $S_{decay}>1$ which results in the constraint, \cite{bl}
\begin{eqnarray}
10^{-40}<\frac{\Gamma_{\sigma}}{M_P}<\left(\frac{4\pi}{3}\right)^2\left(\frac{\sigma_*}{M_P}\right)^4\left(\frac{m_{\sigma}}{M_P}\right).
\end{eqnarray}
This implies $\left(\frac{\sigma_*}{M_P}\right)^4\left(\frac{m_{\sigma}}{M_P}\right)>10^{-40}$ which is  satisfied in the allowed region in parameter space
as can be seen from figures \ref{fig2} and \ref{fig4}.

\section{Conclusions}

In \cite{kk} it was shown that cosmologically relevant magnetic fields can result from gravitationally coupled electrodynamics motivated by the form of the one-loop effective action of the vacuum polarization in QED in a gravitational background. 
Perturbations in the electromagnetic field are amplified during power law inflation determined by the exponent $\beta$ and the value of the Hubble parameter at the end of inflation $H_1$ which here also determines the reheat temperature. The ratio $r$ of magnetic field energy density over background radiation energy density has to be larger than $10^{-37}$, corresponding to $B_s\sim 10^{-20}$ G to seed the galactic dynamo in a universe with no cosmological constant. This value is reduced to $r\sim 10^{-57}$ and $B_s\sim 10^{-30}$ G in a universe with $\Lambda>0$ \cite{dt}.

In \cite{kk} it was found that  there is a region in the $(\beta, \left(\frac{H_1}{M_P}\right))$-plane for which   seed magnetic field are obtained with $B_s>10^{-20}$ G corresponding to $r>10 ^{-37}$. However, the curvature perturbations generated during inflation for those values are incompatible with the observed nearly scale invariant spectrum. Therefore, here the possibility of completing this model of magnetic field generation with the curvaton mechanism has been considered. Thereby generating the curvature perturbations after inflation. Hence imposing that the amplitude of the curvature perturbations during inflation is negligible. This restricts significantly the parameter space in $\beta$ and $\left(\frac{H_1}{M_P}\right)$ allowing the creation of primordial magnetic fields only with amplitude $B_s\sim 10^{-30}$ G, however,  still satisfying the weaker constraint $r>10^{-57}$.

We have assumed that power law inflation is realized within slow roll inflation with an exponential potential.
The curvaton is described by a simple quadratic potential.
Assuming that the curvature perturbation due to the curvaton is determined by the best fit values of WMAP 7 and that the contribution of the curvature perturbation due to the inflaton is less than $10^{-3}$ of that of the curvaton the parameter space of the curvaton model has been explored. The corresponding values in the $(\beta, \left(\frac{H_1}{M_P}\right))$-plane have been found for the mass of the curvaton $m_{\sigma}$, its field value during inflation $\sigma_*$  and the decay constant $\Gamma_{\sigma}$ in the two cases where the curvaton decays during radiation domination or during curvaton domination.

\section{Acknowledgements}
I am indebted to David Lyth for suggesting the problem and for very useful discussions.
Financial support by Spanish Science Ministry grants FPA2009-10612, FIS2009-07238 and CSD2007-00042 is gratefully acknowledged.

\end{document}